\documentclass[prd,nofootinbib,preprint,showpacs,showkeys,superscriptaddress]{revtex4}
\usepackage[english]{babel}
\usepackage{amsmath,amssymb}
\usepackage[dvips]{graphicx}
\usepackage{ifthen,array}
\usepackage[sort&compress]{natbib}
\usepackage{amsfonts}
\usepackage{bbm}

\DeclareMathAlphabet{\mathsc}{OT1}{cmr}{m}{sc}

\allowdisplaybreaks

\def\10{$SO(10)$}
\def\21{SU(2) $\otimes$ U(1) }

\def\422{$SU(4) \otimes SU(2) \otimes SU(2)$}
\def\321{SU(3) $\otimes$ SU(2) $\otimes$ U(1)}

\def\lsim{\raise0.3ex\hbox{$\;<$\kern-0.75em\raise-1.1ex\hbox{$\sim\;$}}}
\def\gsim{\raise0.3ex\hbox{$\;>$\kern-0.75em\raise-1.1ex\hbox{$\sim\;$}}}
\def\vev#1{\left\langle #1\right\rangle}

\newcommand{\AddrAHEP}{%
  AHEP Group, Institut de F\'{\i}sica Corpuscular --
  C.S.I.C./Universitat de Val{\`e}ncia \\
  Edificio Institutos de Paterna, Apt 22085, E--46071 Valencia, Spain}
\newcommand{\AddrLisb}{%
 Departamento de F\'\i sica and CFTP, Instituto Superior T\'ecnico\\
          Av. Rovisco Pais 1, $\:\:$ 1049-001 Lisboa, Portugal }
\begin{document}

\preprint{IFIC/05-63}

\vspace*{2cm} \title{Production and decays of supersymmetric Higgs
  bosons in spontaneously broken R-parity}

\author{M.~Hirsch} \email{mahirsch@ific.uv.es}\affiliation{\AddrAHEP}
\author{J.~C.~Romao}
\email{jorge.romao@ist.utl.pt}\affiliation{\AddrLisb}
\author{J.~W.~F.~Valle} \email{valle@ific.uv.es}
\affiliation{\AddrAHEP} \author{A.~Villanova del Moral}
\email{Albert.Villanova@ific.uv.es}\affiliation{\AddrAHEP}


\begin{abstract}
  We study the mass spectra, production and decay properties of the
  lightest supersymmetric CP-even and CP-odd Higgs bosons in models
  with spontaneously broken R-parity (SBRP).  We compare the resulting
  mass spectra with expectations of the Minimal Supersymmetric
  Standard Model (MSSM), stressing that the model obeys the upper
  bound on the lightest CP-even Higgs boson mass.
  We discuss how the presence of the additional scalar singlet states
  affects the Higgs production cross sections, both for the 
  Bjorken process and the ``associated production''.
  The main phenomenological novelty with respect to the MSSM comes
  from the fact that the spontaneous breaking of lepton number leads
  to the existence of the majoron, denoted $J$, which opens new decay
  channels for supersymmetric Higgs bosons.
  We find that the invisible decays of CP-even Higgses can be
  dominant, while those of the CP-odd bosons may also be sizeable.

\end{abstract}


\maketitle

\section{Introduction
        \label{Introduction}}

      Unveiling the mechanism of symmetry breaking and mass generation
      constitutes one of the main goals in the agenda of upcoming
      accelerators, like CERN's Large Hadron Collider (LHC) and the
      International Linear Collider (ILC).
      Precision electroweak data currently hint that the mechanism
      responsible for electroweak symmetry
      breaking~\cite{Carena:2002es} involves a weakly-coupled Higgs
      sector, as predicted by supersymmetry.
      Supersymmetry also stabilizes the Higgs boson mass against
      quadratic divergences, thus accounting for the hierarchy between
      the electroweak and the Planck scales in a technically natural
      way.
      A very exciting possibility is that the experimentally
      observed~\cite{McDonald:2004dd,Kajita:2004ga,Araki:2004mb}
      neutrino masses and mixings~\cite{Maltoni:2004ei} have a
      supersymmetric origin~\cite{Hirsch:2004he}.
      The key requirement for this to be possible is that R--parity,
      defined as $R_p = (-1)^{3B+L+2S}$ (with $S$, $B$, $L$ denoting
      spin, baryon and lepton numbers, respectively) be violated.
      The simplest way to do this is through a bilinear term in the
      superpotential.

      The resulting model is interesting in two ways.
      First it provides the most economical description of R-parity
      violation as a ``perturbation'' to the Minimal Supersymmetric
      Standard Model: it may be taken as the reference R-parity
      violation model, which we may call RMSSM~\cite{diaz:1998xc}.
      The model offers a minimal low-scale mechanism to generate
      neutrino masses, that successfully accounts for the observed
      pattern of neutrino masses and mixing
      \cite{Hirsch:2000ef,romao:1999up}~\footnote{For
        papers dealing with trilinear terms see, for example
        \cite{chun:1999bq,abada:2001zh}.}.
      In contrast to the seesaw mechanism, it makes well defined
      predictions that will be tested at upcoming colliders LHC/ILC,
      namely, the decay branching ratios of the lightest supersymmetric
      particle are related to the neutrino mixing angles measured in
      neutrino
      experiments~\cite{porod:2000pw}.

      On the other hand the model also provides the simplest effective
      description of theories where the breaking of R-parity occurs
      spontaneously, like that of the electroweak gauge symmetry
      itself, due to the existence of non-zero singlet sneutrino
      vacuum expectation values
      (vevs)~\cite{masiero:1990uj,romao:1992vu,romao:1997xf,shiraishi:1993di}.
      A general feature of models where neutrino masses arise from
      low-scale spontaneous violation of ungauged lepton number is
      that the lightest CP-even supersymmetric Higgs boson will have
      an important decay channel into the singlet Goldstone boson
      (called majoron) associated to lepton number
      violation~\cite{Joshipura:1993hp}:
      \begin{equation}
        \label{eq:HJJ}
        h \to J J\,.
      \end{equation}
      Thus the Higgs boson may decay mainly to an invisible mode
      characterized by missing energy, instead of the Standard Model
      channels.
      This general possibility can also be realized in spontaneously
      broken R--parity supersymmety~\cite{romao:1992zx}.

      We have recently reanalyzed this suggestion in view of the data
      on neutrino oscillations that indicate non-zero neutrino
      masses~\cite{Hirsch:2004rw}. We found that this proposal remains
      valid, despite the smallness of neutrino masses required to fit
      current neutrino oscillation data~\cite{Maltoni:2004ei}.
      In Ref.~\cite{Hirsch:2004rw} we have shown explicitly that the
      invisible decays of the lightest CP-even Higgs boson can be
      dominant, unsuppressed by the small neutrino masses, for the
      same parameter values for which Higgs production in $e^+e^-$
      annihilation is comparable in cross section to that
      characterizing the standard case.
      A necessary ingredient in this case is the existence of an \21
      singlet superfield $\Phi$ coupling to the electroweak doublet
      Higgses, which may provide a solution to the so-called
      $\mu$-problem.  

      In this follow-up paper we extend the analysis and study in
      detail the possibility that the lightest CP-even Higgs boson is
      produced also in association with a CP-odd boson in
      electron-positron collisions.
      The first aspect to consider is the theoretically expected mass
      spectra of CP-even and CP-odd scalar bosons. An important
      feature of any supersymmetric model is the existence of an upper
      bound on the mass of the lightest CP-even scalar boson. We
      verify explicitly that this feature emerges in the present
      model. We also explain how the supersymmetric Higgs boson mass
      upper limit should be understood in terms of the SBRP model
      fields.

      Then we turn to the Higgs production cross sections.
      Although individual Higgs boson production cross sections, via
      the familiar Bjorken process or the associated mode, are
      potentially suppressed with respect to those of the MSSM, given
      enough center-of-mass energy and luminosity, all Higgses can be
      potentially explored due to unitarity.
      We carefully analyze the decay properties of the first and second 
      lightest Higgs
      bosons. The case of the lightest CP-even Higgs boson was
      already considered in Ref.~\cite{Hirsch:2004rw}. We revisit this
      case, and confirm that 
      the invisible dacay mode for either the first or the second
      lightest CP-even
      Higgs boson can easily dominate, in contrast to that of the
      lightest CP-odd 
      which may arise at
      subleading level up to 20 \% level at most.

\section{The Model \label{The Model}}

For completeness we recall here the main ingredients of the model. In
addition to the Minimal Supersymmetric Standard Model superfields it
contains \21 singlet superfields $({\widehat \nu^c}_i,\widehat{S}_i,
{\widehat \Phi})$ carrying lepton number assigned as $(-1, 1,0)$. With
this choice the most general superpotential terms conserving lepton
number are given as~\cite{romao:1992zx}
\begin{eqnarray} \nonumber
{\cal W} &\hskip-4mm=\hskip-4mm& \varepsilon_{ab}\Big(
h_U^{ij}\widehat Q_i^a\widehat U_j\widehat H_u^b 
+h_D^{ij}\widehat Q_i^b\widehat D_j\widehat H_d^a 
+h_E^{ij}\widehat L_i^b\widehat E_j\widehat H_d^a 
+h_{\nu}^{ij}\widehat L_i^a\widehat \nu^c_j\widehat H_u^b 
\!- {\hat \mu}\widehat H_d^a\widehat H_u^b 
\!- (h_0 \widehat H_d^a\widehat H_u^b +\delta^2)\widehat\Phi \Big)  \\
& &+\hskip 5mm   h^{ij} \widehat\Phi \widehat\nu^c_i\widehat S_j +
M_{R}^{ij}\widehat \nu^c_i\widehat S_j 
+ \frac{1}{2}M_{\Phi} \widehat\Phi^2 +\frac{\lambda}{3!} \widehat\Phi^3\,.
\label{eq:Wsuppot} 
\end{eqnarray}
The first three terms together with the $ {\hat \mu}$ term define the
R-parity conserving MSSM, the terms in the last row involve only
couplings among the \21 singlet superfields $({\widehat
  \nu^c}_i,\widehat{S}_i,{\widehat \Phi})$. The remaining terms couple
the singlets to the MSSM fields. We stress the importance of the
Dirac-Yukawa term which connects the right-handed neutrino superfields
to the lepton doublet superfields, thus fixing lepton number.

Like all other Yukawa couplings in general $h_{\nu}$ is an arbitrary 
non-symmetric complex matrix in generation space. However, for technical 
simplicity we will consider only the case with just one pair 
of lepton--number--carrying \21 singlet superfields, $\widehat\nu^c$ 
and $\widehat S$, in order to avoid inessential complication. This in 
turn implies, $ h^{ij} \to h$ and $ h_{\nu}^{ij} \to h_{\nu}^{i}$. 

The scalar potential along neutral directions is given by
\begin{eqnarray}
V_{total}  &=&
|h \Phi \tilde{S} + h_{\nu}^{i} \tilde{\nu}_i H_u  + M_R \tilde{S}|^2 +
|h_0 \Phi H_u + \hat{\mu} H_u|^2 + |h \Phi \tilde{\nu^c}+ M_R
\tilde{\nu^c}|^2 
\label{scalarpot}
\\\nonumber&&
+|- h_0 \Phi H_d  - \hat{\mu} H_d +
h_{\nu}^{i} \tilde{\nu}_i \tilde{\nu^c} |^2+
|- h_0 H_u H_d + h \tilde{\nu^c} \tilde{S} - \delta^2 +M_{\Phi} \Phi 
+\frac{\lambda}{2} \Phi^2|^2 \\\nonumber&&
+\sum_{i=1}^3|h_{\nu}^{i} \tilde{\nu^c} H_u|^2
+ \Big[  A_h h \Phi \tilde{\nu^c} \tilde{S}
- A_{h_0} h_0 \Phi H_u H_d 
+ A_{h_{\nu }}  h_{\nu}^{i}\tilde{\nu}_i H_u \tilde{\nu^c} 
- B \hat{\mu} H_u H_d  \\\nonumber&&
- C_{\delta} \delta^2 \Phi + B_{M_R} M_R \tilde{\nu^c} \tilde{S} 
+ \frac{1}{2} B_{M_{\Phi}} M_{\Phi} \Phi^2 
+ \frac{1}{3!} A_{\lambda} \lambda \Phi^3 
 + h.c. \Big]\\\nonumber&&
+ \sum_{\alpha} \tilde{m}_{\alpha}^2 |z_{\alpha}|^2
+ \frac{1}{8} (g^2 + {g'}^2) 
\Big( |H_u|^2 - |H_d|^2 - \sum_{i=1}^3 |\tilde{\nu}_i|^2\Big)^2\,,
\end{eqnarray}
where $z_{\alpha}$ denotes any neutral scalar field in the theory.
For simplicity we assume CP conservation in the scalar sector,
taking all couplings real.

Electroweak symmetry breaking is driven by the isodoublet vevs
$\vev{{H_u}} = \frac{v_{u}}{\sqrt{2}}$ and $\vev{{H_d}}=
\frac{v_{d}}{\sqrt{2}}$, with the combination $v^2 = v_u^2 + v_d^2 +
\sum_i v_{L i}^2$ fixed by the $W$ mass, while the ratio of isodoublet
vevs yields $\tan \beta = \frac{v_u}{v_d}$.  Here,
$\frac{v_{Li}}{\sqrt{2}}$ are the vevs of the left-scalar neutrinos.
They vanish in the limit where $h^i_{\nu} \to 0$. In this limit
R--parity is restored and neutrinos become massless, as in the MSSM,
and, apart from $\Phi$, the extra singlets become phenomenologically
irrelevant, one reaches the NMSSM
limit~\cite{Panagiotakopoulos:2000wp,Dedes:2000jp}.

The spontaneous breaking of R-parity is driven by nonzero vevs for the
right-scalar neutrinos. The scale characterizing R-parity breaking is
set by the isosinglet vevs $\vev{\tilde{\nu^c}} =
\frac{v_R}{\sqrt{2}}$ and
$\langle\tilde{S}\rangle=\frac{v_S}{\sqrt{2}}$. Finally, $\vev{\Phi} =
\frac{v_{\Phi}}{\sqrt{2}}$ gives a contribution to the $\mu$--term.

With the above choices and definitions we can obtain the neutral scalar 
boson mass matrices as described in Ref.~\cite{romao:1992vu}. This results in 
$8 \times 8$ mass matrices for the real and imaginary parts of the 
neutral scalars. Their complete definition can be found 
in ~\cite{Hirsch:2004rw}. The spontaneous breaking of \21 and 
lepton number leads to two Goldstone bosons, namely $G^0$ , the one 
``eaten'' by the $Z^0$, as well as $J$, the majoron.
In the basis
$P'^0=(H_d^{0 I},H_u^{0 I},\tilde{\nu}^{1 I},\tilde{\nu}^{2 I},
\tilde{\nu}^{3 I},\Phi^I,\tilde{S}^I,\tilde{\nu}^{c I})$ 
these fields are given as,
\begin{eqnarray}
  \label{eq:3a}
  G^0&=&(N_0\, v_d,-N_0\, v_u,N_0\,v_{L1},N_0\, v_{L2},N_0\, v_{L3},0,0,0)\,
  \nonumber \\[+2mm]
  J&=&N_4 (-N_1 v_d,N_1 v_u, N_2 v_{L1}, N_2 v_{L2}, N_2 v_{L3},0,
  N_3 v_S,-N_3 v_R)\,,
\end{eqnarray}
where the normalization constants $N_i$ are given as
\begin{eqnarray}
  \label{eq:4a}
  N_0&=&\frac{1}{\sqrt{v_d^2+v_u^2+v_{L1}^2 + v_{L2}^2 + v_{L3}^2}}
\nonumber \\[+2mm]
  N_1&=&v_{L1}^2 + v_{L2}^2 + v_{L3}^2\nonumber \\[+2mm]
  N_2&=&v_d^2 + v_u^2\nonumber \\[+2mm]
  N_3&=&N_1 + N_2 \nonumber \\[+2mm]
  N_4&=&\frac{1}{\sqrt{N_1^2 N_2 + N_2^2 N_1 +N_3^2(v_R^2+v_S^2)}}
\end{eqnarray}
and can easily be checked to be orthogonal, i.~e. they satisfy $G^0
\cdot J=0$. 

The neutrino masses and mixings arising from this
model~\cite{Hirsch:2004rw} have been shown to reproduce the current
data on neutrino oscillations that indicate non-zero neutrino
masses~\cite{Maltoni:2004ei}.
Since neutrino masses are so much smaller than all other fermion mass
terms in the model, once can find the effective neutrino mass matrix
in a seesaw--type approximation. After some algebraic manipulation, 
the effective neutrino mass matrix can be cast into a very simple form 
\begin{equation}
(\boldsymbol{m_{\nu\nu}^{\rm eff}})_{ij} = a \Lambda_i \Lambda_j + 
     b (\epsilon_i \Lambda_j + \epsilon_j \Lambda_i) +
     c \epsilon_i \epsilon_j\,,
\label{eq:eff}
\end{equation}
where one can define the effective bilinear R--parity violating
parameters $\epsilon_{i}$ and $\Lambda_i$ as
\begin{equation}
  \label{eq:eps}
\epsilon_{i} = h_{\nu}^{i}\, \frac{v_R}{\sqrt{2}}  
\end{equation}
and
\begin{equation}
\Lambda_i = \epsilon_i v_d + \mu v_{L_i}\,.
\label{eq:deflam0}
\end{equation}
Here the parameter $\mu$ is
\begin{equation}
\mu = \hat{\mu} + h_0 \frac{v_{\Phi}}{\sqrt{2}}\,,
\label{eq:defmu}
\end{equation}
while the coefficients appearing in Eq.~(\ref{eq:eff}) are given in
Ref.~\cite{Hirsch:2004rw}. Eq.~(\ref{eq:eff}) resembles closely the
structure found for the explicit bilinear model at the one-loop level.
However, the coefficients are different, see \cite{Hirsch:2004rw}.

Neutrino physics puts a number of constraints on the parameters 
$\Lambda_i$ and $\epsilon_i$. For the current paper, however, 
exact details are unimportant, the most essential constraint 
for the following discussion is that $h_{\nu}^i \ll 1$ is 
required. (See later discussion of left-sneutrino mixing. In 
the limit $h_{\nu}^i=0$ left-sneutrinos do not mix at all 
with Higgses and singlets).

The requirement that $v_{L_i} \ll v$ can be used to find a 
simple approximation formula for the majoron, given by 
\begin{equation}
\label{eq:majntl}
J \simeq  (\frac{-v_d v_L^2}{Vv^2},\frac{v_u v_L^2}{Vv^2},
           \frac{v_{L1}}{V},\frac{v_{L2}}{V},\frac{v_{L3}}{V},
           0,\frac{v_S}{V},-\frac{v_R}{V})\,,
\end{equation}
where $ V^2=v_S^2+v_R^2$. Thus, the majoron is essentially made up of
the ${\tilde \nu}^c$ and ${\tilde S}$ fields.  This will be important
later, when we discuss the decays of the Higgs bosons.

\section{Higgs spectrum}
\label{sec:higgs-spectrum}

Let us first briefly discuss the spectrum of the scalar and
pseudo-scalar sectors in the model. For detailed definitions we refer
the reader to Ref.~\cite{Hirsch:2004rw}. Since these mass matrices are
too complicated for analytic diagonalization, we will solve the exact
eigensystems numerically. However, before doing that, we discuss
certain limits, where some simplifying approximations are made.  This
allows us to gain some insight into the nature of the spectra.

In the SBRP model there are 8 neutral CP-even states $S^0_i$. In the
neutral CP-odd sector there are six massive states $P^0_i$ ($i=1,\ldots,6$), in addition
to the majoron $J$, with $m_J=0$, and the Goldstone $G^0$.  We
introduce the convention, to be discussed below:
\begin{eqnarray}
\label{eq:defstates}
\left(S^0\right)^T &=&\left(
 S_{h^0},S_{H^0},S_J,S_{J_\perp},S_{\Phi},S_{\tilde\nu_i} \right)
\\ \nonumber
\left(P^0\right)^T &=&\left(
P_{A^0},P_{J_\perp},P_{\Phi},P_{\tilde\nu_i},J, G^0\right).
\end{eqnarray}
Note, that the ordering of these states is not by increasing mass,
as we have defined $P^0_i$ ($i=1,\ldots,6$) as the massive states.

First we note that all entries in the sub-matrices which mix the
left-sneutrinos to the doublet Higgses and the singlet states are
proportional to $h^i_{\nu}$. In the region of parameters where the
model accounts for the observed neutrino masses we must have that
$\epsilon_i = h^i_{\nu}v_R/\sqrt{2}$ is necessarily a small number and
therefore $h^i_{\nu} \ll 1$. Thus, left sneutrinos mix very little
with the other (pseudo-)scalars, unless entries in the sneutrino
sector are, by chance, highly degenerate with the ones in the other
sectors. The real (imaginary) parts of these nearly-sneutrino states
are denoted by $S_{\tilde\nu_i}$ ($P_{\tilde\nu_i}$) in the definition
given above.  Barring fine-tuned situations, we conclude that mixing
between Higgses and left sneutrinos will, in general, be small.

Consider now the pseudoscalar sector,
\begin{eqnarray}
  \label{eq:mp2}
\boldsymbol{M^{P^2}} = \left[
    \begin{array}{lll}
      \boldsymbol{M^{P^2}_{HH}} & \boldsymbol{M^{P^2}_{H\widetilde L}} & \boldsymbol{M^{P^2}_{HS}}\\[+2mm]
      \boldsymbol{M^{P^2}_{H\widetilde L}}\!{}^{\rm T} & \boldsymbol{M^{P^2}_{\widetilde L \widetilde L}} 
      & \boldsymbol{M^{P^2}_{\widetilde L S}}\\[+2mm]
      \boldsymbol{M^{P^2}_{HS}}\!{}^{\rm T} &\boldsymbol{M^{P^2}_{\widetilde L S}}\!{}^{\rm T} & \boldsymbol{M^{P^2}_{SS}}
    \end{array}
    \right],
\end{eqnarray}
where $\boldsymbol{M^{P^2}_{HH}}$ is a symmetric $2\times 2$ matrix, $\boldsymbol{M^{P^2}_{\widetilde L
\widetilde L}} $ and $\boldsymbol{M^{P^2}_{SS}} $ are symmetric $3\times 3 $ matrices,
while   $\boldsymbol{M^{P^2}_{H\widetilde L}}$ and $\boldsymbol{M^{P^2}_{HS}}$ are $2\times 3$ matrices
and finally $M^{P^2}_{\widetilde L S}$ is (a non-symmetric) $3\times 3$
matrix. In this notation $\widetilde L$ denotes the sneutrinos and $S$
the singlet fields. 

Neglecting terms proportional to $h^i_{\nu}$, $\boldsymbol{M^{P^2}_{HH}}$ can be 
written as
\begin{eqnarray}
  \label{eq:mphh}
\boldsymbol{M^{P^2}_{HH}} = \left[
    \begin{array}{ll}
     \Omega \frac{v_u}{v_d}& \Omega \\
     \Omega & \Omega \frac{v_d}{v_u}
    \end{array}
    \right],
\end{eqnarray}
where~\footnote{We correct a misprint in Ref.~\cite{Hirsch:2004rw}.}
\begin{equation}
  \label{eq:omega}
   \Omega= B \hat\mu
   -\delta^2 h_0 + \frac{\lambda}{4} h_0 v_{\Phi}^2+\frac{1}{2} h h_0
   v_R v_S + \frac{\sqrt{2}}{2} A_{h_0} h_0 v_{\Phi} +
  \frac{\sqrt{2}}{2} h_0 M_{\Phi} v_{\Phi}\,.
\end{equation}
Note the presence of $h_0$-dependent terms in Eq.~(\ref{eq:omega}). If
there were no mixing between the doublet and singlet Higgses,
Eq.~(\ref{eq:mphh}) would yield the eigenvalues,
\begin{eqnarray}
  \label{eq:mphheigs}
m^2_{1,2} = \Big( 0, \Omega(\frac{v_u}{v_d}+\frac{v_d}{v_u})\Big)\,,
\end{eqnarray}
with the massless state identified as the Goldstone boson, $G^0$, 
and the other state as the pseudo-scalar Higgs $A^0$ of the 
MSSM, with 
\begin{equation}
  \label{eq:massAD}
  m^2_{A^0}=\frac{2 \Omega}{\sin 2\beta}\,.
\end{equation}
The state most closely resembling the MSSM $A^0$, i.e. the state 
remaining in the spectrum when singlets are decoupled is called 
$P_{A^0}$ in Eq.~(\ref{eq:defstates}).
The sub-matrix $\boldsymbol{M^{P^2}_{SS}}$, on the other hand, in the limit 
$h^i_{\nu}=0$, can be written as,
\begin{eqnarray}
  \label{eq:mpss}
\boldsymbol{M^{P^2}_{SS}} = \left[
    \begin{array}{lll}
     M^{P^2}_{SS_{11}} & M^{P^2}_{SS_{12}} & M^{P^2}_{SS_{13}}\\
     M^{P^2}_{SS_{12}} & - \Gamma \frac{v_R}{v_S}& - \Gamma \\
     M^{P^2}_{SS_{13}} & - \Gamma & - \Gamma \frac{v_S}{v_R}
    \end{array}
    \right],
\end{eqnarray}
where,
\begin{eqnarray}
  \label{eq:mpssdef}
M^{P^2}_{SS_{11}} &=& \delta^2  \left(
    C_{\delta} + M_{\Phi} \right) \frac{\sqrt{2}}{v_{\Phi}}
  -\frac{\sqrt{2}}{2} (v_d^2 + v_u^2) \frac{h_0 \hat\mu}{v_{\Phi}} -
\frac{\sqrt{2}}{4} \lambda \left( 3 A_{\lambda} + M_{\Phi}\right)
    v_{\Phi} \nonumber \\
  & - & 2 B_{M_{\Phi}} M_{\Phi} 
- \frac{\sqrt{2}}{2} h \left(A_h +  M_{\Phi} \right)
    \frac{v_R v_S}{v_{\Phi}} 
+ \frac{\sqrt{2}}{2} h_0 \left(A_{h_0} + M_{\Phi}\right)
\frac{v_u v_d }{v_{\Phi}} \nonumber \\
&+& 2 \delta^2 \lambda + \lambda h_0\, v_u v_d -
\lambda h\, v_R v_S - \frac{\sqrt{2}}{2} h\, M_R
\frac{v_S^2+v_R^2}{v_{\Phi}}\, , 
\\[+2mm]
M^{P^2}_{SS_{12}} &=& -\frac{1}{\sqrt{2}} h(A_h-{\hat M}_{\Phi}) v_R\,, \\
M^{P^2}_{SS_{13}} &=& -\frac{1}{\sqrt{2}} h(A_h-{\hat M}_{\Phi}) v_S\,. 
\end{eqnarray}
Here ${\hat M}_{\Phi}= M_{\Phi} + \lambda v_\Phi/\sqrt2$ and
\begin{equation}
  \label{eq:gamma}
  \Gamma=B_{M_R} M_R -\delta^2 h + \frac{1}{4} h \lambda v_{\Phi}^2 -
  \frac{1}{2} h h_0 v_u v_d + \frac{\sqrt{2}}{2} h \left( A_h + 
  M_{\Phi}\right) v_{\Phi}\, .
\end{equation}
Eq.~(\ref{eq:mpss}) has one zero eigenvalue, approximately identified 
with the majoron, $J$, and two non-zero eigenvalues. If 
$M^{P^2}_{SS_{12}},M^{P^2}_{SS_{13}} \ll M^{P^2}_{SS_{11}}+ \Gamma$ 
then the eigenvalues of Eq.~(\ref{eq:mpss}) are approximately given by
\begin{eqnarray}
\label{eq:mpsseigs}\nonumber
m^2_{1,2,3} = \Big( 0, &-& \Gamma(\frac{v_R}{v_S}+\frac{v_S}{v_R})-
\frac{1}{2} \frac{h^2(A_h-{\hat M}_{\Phi})^2 v_R^2v_S^2}
                 {M^{P^2}_{SS_{11}}v_R v_S+\Gamma(v_R^2+v_S^2)} + \cdots, \\
& &M^{P^2}_{SS_{11}}
+\frac{1}{2} \frac{h^2(A_h-{\hat M}_{\Phi})^2v_R v_S(v_R^2+v_S^2)}
                 {M^{P^2}_{SS_{11}}v_R v_S+\Gamma(v_R^2+v_S^2)} + \cdots
\Big)\,,
\end{eqnarray}
where the dots stand for higher order terms. The eigenvalue
proportional to $\Gamma$ is mainly a combination of
$\tilde{S}^I,\tilde{\nu}^{c I}$ fields and we call it $P_{J_\perp}$ in
Eq.~(\ref{eq:defstates}) above, because in the limit where $m_{P_\Phi}
\rightarrow \infty$ and $v_{L_i} \to 0$ this massive state is
orthogonal to the majoron.  As we will discuss below, it is this state
which preferably decays invisibly. The third eigenvalue in Eq.~(\ref{eq:mpsseigs}) 
is an approximation to the state called $P_{\Phi}$
above.  Due to mixing between doublet and singlet states both Eq.~(\ref{eq:mphheigs}) 
and Eq.~(\ref{eq:mpsseigs}), are only very crude estimates.

Consider the scalar sector of the model,
\begin{eqnarray}
  \label{eq:1a}
  \boldsymbol{M^{S^2}}=\left[
    \begin{array}{lll}
      \boldsymbol{M^{S^2}_{HH}} & \boldsymbol{M^{S^2}_{H\widetilde L}} & \boldsymbol{M^{S^2}_{HS}}\\[+2mm]
      \boldsymbol{M^{S^2}_{H\widetilde L}}\!{}^{\rm T} & \boldsymbol{M^{S^2}_{\widetilde L \widetilde L}} 
      & \boldsymbol{M^{S^2}_{\widetilde L S}}\\[+2mm]
      \boldsymbol{M^{S^2}_{HS}}\!{}^{\rm T} &\boldsymbol{M^{S^2}_{\widetilde L S}}\!{}^{\rm T} & \boldsymbol{M^{S^2}_{SS}}
    \end{array}
    \right],
\end{eqnarray}
where the blocks have the same structure as before. $\boldsymbol{M^{S^2}_{HH}}$ 
contains two eigenvalues which, in the limit of zero mixing, would 
be identified with the MSSM states $h^0$ and $H^0$. \footnote{As 
in the MSSM, there is an upper limit for the mass of the $S_{h^0}$, see 
the discussion below.} These states are the ones called $S_{h^0}$ 
and $S_{H^0}$ in Eq.~(\ref{eq:defstates}) above. 

The sub-matrix $\boldsymbol{M^{S^2}_{SS}}$ contains, in general, three non-zero 
eigenvalues. One can find an approximate analytic expression for 
them in the limit that the state $S_\Phi$ is much heavier than the 
remaining two eigenstates (called $S_J$ and $S_{J_\perp}$). Again 
in the limit of small mixing, the eigenvalues of the latter are 
approximately given by
\begin{eqnarray}
  \label{eq:mssseigs}
m^2_{1,2} = \Big(2 h^2\frac{v_R^2 v_S^2}{(v_R^2+v_S^2)}+ \cdots,
           -\Gamma(\frac{v_R}{v_S}+\frac{v_S}{v_R})-
2 h^2\frac{v_R^2 v_S^2}{(v_R^2+v_S^2)}+\cdots\Big)
\end{eqnarray}
The first (second) of the eigenvalues in Eq.~(\ref{eq:mssseigs}) is
approximately the state $S_J$ ($S_{J_\perp}$).

\begin{figure}[htbp]
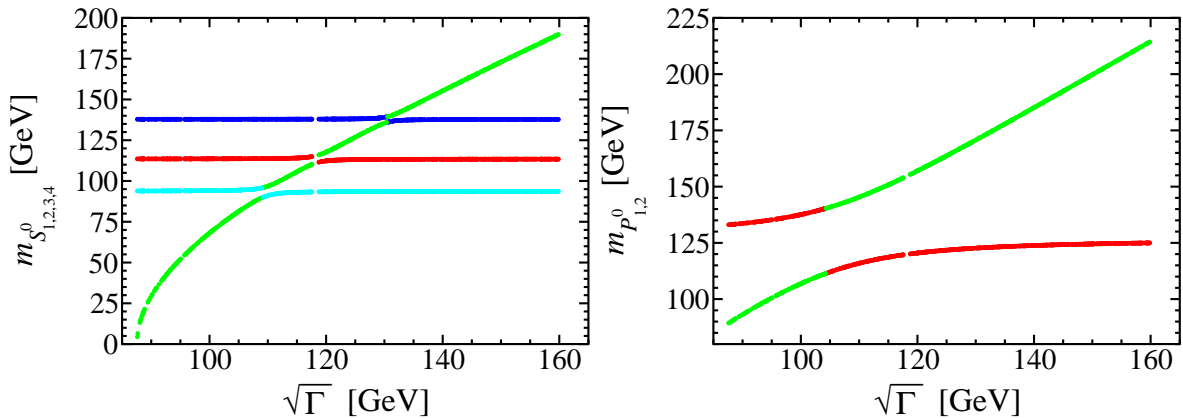

  \centering
  \includegraphics[clip,width=0.47\linewidth]{mH-Gamma.eps}
  \includegraphics[clip,width=0.47\linewidth]{mA-Gamma.eps}
  \caption{Typical CP-even (left) and CP-odd (right) Higgs masses as
    function of the parameter $\Gamma$. In this example there are four
    light CP-even states and two light massive CP-odd states (plus two
    massless states, $G^0$ and $J$, not shown). Just as in the MSSM
    there is always one light doublet state, coinciding with $h^0$ in
    the limit of zero mixing. Other states can (but need not) be
    light, depending on the parameters $\Omega$ and $\Gamma$, see
    text.}
  \label{fig:masses}
\end{figure}

Fig.~\ref{fig:masses}, to the left, shows an example of the four
lowest lying eigenvalues in the CP-even sector, as a function of
$\Gamma$ for a random but fixed choice of the remaining parameters.
One of the states, $S_{J_\perp}$, which is mainly singlet, is proportional to
$\Gamma$, as indicated by Eq.~(\ref{eq:mssseigs}). There is another
singlet state, corresponding to $S_J$ of Eq.~(\ref{eq:mssseigs}), and
two mainly doublet states, identified with $S_{h^0}$ and $S_{H^0}$. We
note in passing that $m_{S_{H^0}}$ is proportional to $\Omega$, as in
the MSSM. Mixing between singlet and doublet states will be important
always if the eigenvalues are comparable, as for the example shown in
the figure.  Thus, all the discussion above should be taken as
qualitative only.

The right panel in Fig.~\ref{fig:masses} shows an example of the
two lightest massive CP-odd eigenvalues as a function of $\Gamma$ for a fixed
but random set of other parameters. That one eigenvalue is
proportional to $\Gamma$ is obvious from Eq.~(\ref{eq:mpsseigs}). We note that $\Omega$ and
$\Gamma$ are the main parameters which will determine associated
production and influence the branching ratio into invisible states, as
we will discuss in the following sections.
The model clearly exhibits decoupling, just as the MSSM. In the limit
where $\Omega$ goes to infinity the masses of both states 
$P_{A^0}$ and $S_{H^0}$ go to infinity, just as what happens in the
MSSM when $m_{A^0}$ goes to infinity. The states $S_{J_\perp}$ and
$P_{J_\perp}$ are decoupled in the limit as $\Gamma$ goes to infinity.
If, in addition, we require $h\ll 1$ also $S_{J}$ decouples and the SM
Higgs phenomenology is recovered, as in the MSSM.

\section{Higgs boson production}
\label{sec:higgs-prod}

Supersymmetric Higgs bosons can be produced at an $e^+ e^-$ collider
through their couplings to $Z^0$, via the so--called Bjorken process
($e^+ e^- \rightarrow Z^0 S^0_i$), or via the associated production mechanism 
($e^+ e^- \rightarrow S^0_i P^0_j$).
In our SBRP model there are 8 neutral CP-even states $S^0_i$ and 6
massive neutral CP-odd Higgs bosons $P^0_i$, in addition to the majoron $J$ and
the Goldstone $G^0$, see Eq.~(\ref{eq:defstates}).

One must diagonalize the (pseudo-)scalar boson mass matrices in order to find the
couplings of the scalars to the $Z^0$. After doing that we obtain
the Lagrangian terms
\begin{equation}
\label{HZZ1}
{\cal L}
\supset\sum_{i=1}^8 (\sqrt 2 G_F)^{1/2} M_Z^2 Z^0_{\mu}Z^{0\mu}\, \eta_{{\rm B}_i} S^0_i
+\sum_{i,j=1}^8(\sqrt{2}G_F)^{1/2}M_Z\, \eta_{{\rm A}_{ij}}
\left(Z^{0\mu}S^0_i\overleftrightarrow{\partial_{\mu}} P^0_j\right)
\end{equation}
with each $\eta_{{\rm B}_i}$ given as a weighted combination of the five \21
doublet scalars,
\begin{equation}
  \label{eq:etaB}
  \eta_{{\rm B}_i}= \frac{v_d}{v} R^{S^0}_{i 1} + \frac{v_u}{v} R^{S^0}_{i 2} + 
\sum_{j=1}^3 \frac{v_{Lj}}{v} R^{S^0}_{i j+2} 
\end{equation}
and the $\eta_{{\rm A}_{ij}}$ given by
\begin{equation}
  \label{eq:etaA}
\eta_{{\rm A}_{ij}}=R^{S^0}_{i1}R^{P^0}_{j1}-R^{S^0}_{i2}R^{P^0}_{j2}+\sum_{k=1}^3
R^{S^0}_{ik+2}R^{P^0}_{jk+2} 
\end{equation}
where the subscripts ${\rm B}$ and ${\rm A}$ refer, respectively, to the Bjorken
process or associated production mechanisms. From these Lagrangian terms we
can easily derive the production cross sections. These are simple
generalizations of the MSSM
results~\cite{Accomando:1997wt,pocsik:1981bg} and for completeness we
give them in Appendix~\ref{sec:prod-cross-sect}.

In the MSSM, there are two sum rule rules, one concerning only the CP
even sector
\begin{equation}
  \label{eq:sumrule1}
  \eta_{{\rm B}_{h^0}}^2 +\eta_{{\rm B}_{H^0}}^2 = 1\,,
\end{equation}
and another relating the Bjorken and the associated production mechanisms,
\begin{equation}
  \label{eq:sumrule2}
  \eta_{{\rm B}_{h^0}}^2 +\eta_{{\rm A}_{h^0A^0}}^2 = 1\,,
\end{equation}
with $\eta_{{\rm B}_{h^0}}=\sin(\alpha-\beta)$ and $\eta_{{\rm A}_{h^0A^0}}=
\eta_{{\rm B}_{H^0}}= \cos(\alpha-\beta)$, in an obvious notation.

It is very easy, and instructive, to use our expressions for $\eta_{\rm A}$
and $\eta_{\rm B}$ to recover the MSSM result in the limit that we have only 
the $H_d$ and $H_u$ doublets. In fact, in this case
\begin{equation}
  \label{eq:3}
  \frac{v_d}{v}=R^{P^0}_{22}\,,\qquad \frac{v_u}{v}=R^{P^0}_{21}\,,
\end{equation}
so we have
\begin{equation}
  \label{eq:1}
  \eta_{{\rm B}_{h^0}}=R^{P^0}_{22} R^{S^0}_{11} + R^{P^0}_{21} R^{S^0}_{12}\,, \quad
  \eta_{{\rm B}_{H^0}}=R^{P^0}_{22} R^{S^0}_{21} + R^{P^0}_{21} R^{S^0}_{22}\,, \quad
  \eta_{{\rm A}_{h^0A^0}}=R^{P^0}_{21} R^{S^0}_{11} - R^{P^0}_{22} R^{S^0}_{12}
\end{equation}
and we get for the sum rule of Eq.~(\ref{eq:sumrule2})
\begin{eqnarray}
  \label{eq:2}
  \eta_{{\rm A}_{h^0A^0}}^2 + \eta_{{\rm B}_{h^0}}^2 &=& 
  \left(R^{P^0}_{22} R^{S^0}_{11} + R^{P^0}_{21} R^{S^0}_{12}\right)^2
  +\left(R^{P^0}_{21} R^{S^0}_{11} - R^{P^0}_{22} R^{S^0}_{12}\right)^2\\
&=& \left( R^{S^0}_{11} R^{S^0}_{11} + R^{S^0}_{12} R^{S^0}_{12} \right)
    \left( R^{P^0}_{21} R^{P^0}_{21} + R^{P^0}_{22} R^{P^0}_{22}\right)\\
&=&1\,,
\end{eqnarray}
where we have used the orthogonality of the rotation matrices
\begin{equation}
  \label{eq:4}
  \sum_{k=1}^2 R^{S^0}_{ik} R^{S^0}_{jk}=\delta_{ij}\,,\quad
  \sum_{k=1}^2 R^{P^0}_{ik} R^{P^0}_{jk}=\delta_{ij}\,, \quad (i,j=1,2)\,.
\end{equation}
For the sum rule of the CP-even sector, Eq.~(\ref{eq:sumrule1}), we
get 
\begin{eqnarray}
  \label{eq:6}
  \eta_{{\rm B}_{h^0}}^2 + \eta_{{\rm B}_{H^0}}^2 &=& 
  \cos^2 \beta \left(R^{S^0}_{11} R^{S^0}_{11} + R^{S^0}_{21}
  R^{S^0}_{21}\right)
 + \sin^2 \beta \left(R^{S^0}_{12} R^{S^0}_{12} + R^{S^0}_{22}
  R^{S^0}_{22}\right)\\
&&+2 \sin\beta \cos\beta \left(
  R^{S^0}_{11} R^{S^0}_{12}+R^{S^0}_{21} R^{S^0}_{22}\right)\\
  &=&1\,,
\end{eqnarray}
using the result that in an orthogonal matrix also the vectors
corresponding to the columns are orthonormal, that is
\begin{equation}
  \label{eq:7}
    \sum_{k=1}^2 R^{S^0}_{ki} R^{S^0}_{kj}=\delta_{ij}\,,\quad (i,j=1,2)\,.
\end{equation}

\begin{figure}[ht]
  \centering
  \includegraphics[clip,height=70mm]{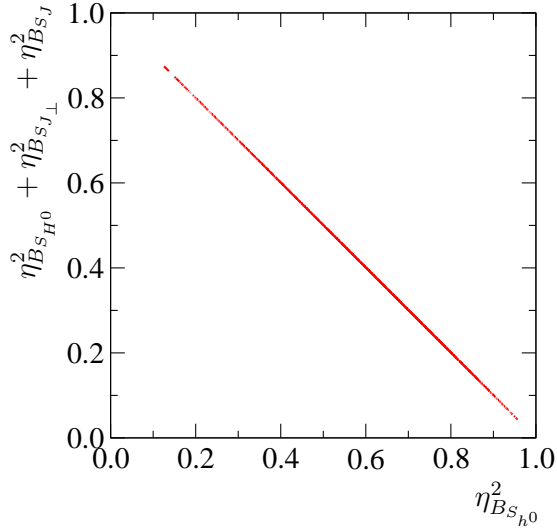} 
  \caption{Sum rule in the CP-even sector, for the case explained in
    the text. The four states, $(S_{h^{0}}, S_{H^{0}}, S_{J_{\perp}},
    S_{J})$. For this example all scalar masses are taken below $200$ GeV.}
  \label{fig:sumrule}
\end{figure}

How this differs in our case? The difference is that, in general, 
\begin{equation}
  \label{eq:5}
  \sum_{k=1}^2 R^{S^0,P^0}_{ik} R^{S^0,P^0}_{jk} \not= \delta_{ij} \quad
  \hbox{and}\quad
  \sum_{k=1}^2 R^{S^0,P^0}_{ki} R^{S^0,P^0}_{kj} \not= \delta_{ij}\,, 
\end{equation}
due to the fact that we now have more than two (pseudo-)scalars. As it was
stated in the last section and will be discussed in more detail when
we consider the decays, to have a sizeable invisible branching ratio
we need the doublets to be close in mass to the singlet states related
to the majoron and orthogonal combinations. This means that, in the CP-even 
sector, the first four states are $(S_{h^{0}}, S_{H^{0}},
S_{J_{\perp}}, S_{J})$, while in the CP-odd sector we should have 
$(P_{A_{0}},P_{J_{\perp}}, J, G^0)$.  If this situation happens then we can
very easily find a generalization of the sum rule of the CP-even
sector, as
\begin{equation}
  \label{eq:8}
  \eta_{{\rm B}_{S_{h^{0}}}}^2 + \eta_{{\rm B}_{S_{H^{0}}}}^2 +
  \eta_{{\rm B}_{S_{J_{\perp}}}}^2+\eta_{{\rm B}_{S_{J}}}^2 =1
\end{equation}
to a good approximation.  This is displayed in Fig.~\ref{fig:sumrule}
where we plot the sum $\eta_{{\rm B}_{S_{H^{0}}}}^2 +
\eta_{{\rm B}_{S_{J_{\perp}}}}^2+\eta_{{\rm B}_{S_{J}}}^2$ against
$\eta_{{\rm B}_{S_{h^{0}}}}^2$.
The significance of this sum rule should be clear: if the lightest
Higgs boson has a very small coupling to the $Z^0$ and hence a small
production cross section, there should be another state nearby that
has a large production cross section. 

The other sum rule, relating the CP-even and CP-odd sectors,
Eq.~(\ref{eq:sumrule2}), is more difficult to generalize. In fact the
$P_{A_{0}}$ state will now mix with the $P_{J_{\perp}}$ and the
identification of Eq.~(\ref{eq:3}) will be no longer true. However
qualitatively the sum rule still holds in the sense that if the
parameters are such that the production of the CP-odd states is
reduced one always gets a CP-even state produced.

The above discussion has concentrated on Higgs boson production at an
$e^+e^-$ collider. We now briefly comment on the differences with
regards to Higgs production at the LHC~\cite{Romao:1992dc}. It has
been suggested to search for an invisibly decaying Higgs at the LHC in
$WW$ boson fusion \cite{Eboli:2000ze}, in asociated production with a $Z^0$
boson \cite{Godbole:2003it}, or in the $t{\bar t}$ channel
\cite{Gunion:1993jf}. For the production in $WW$ fusion or in asociated
production with a $Z^0$ boson the above discussion applies
straightforwardly, since the relevant coupling in both cases is
$\eta_{B_i}$ (i.e. $sin(\beta-\alpha)$ in the MSSM limit). For the
$t{\bar t}$ channel in the MSSM production cross section the factor
$\cos\alpha$ has to be replaced by $R^{S^0}_{i2}$ for the SBRP model.

\section{Higgs boson decays}
\label{sec:higgs-decays}

In the following we will discuss the decays of light CP-even and CP-odd 
supersymmetric Higgs bosons. Since the phenomenology of Higgs
bosons within the MSSM is well-known~\cite{hunters,gunion:1986yn}, we
will concentrate on non-standard final states. Of these, the most
important are the majoron Higgs boson decay modes, which are
characteristic of the SBRP model, without an MSSM counterpart.
We will limit ourselves to the discussion of light states, i.e.  Higgs
bosons with masses below the $2W$ threshold. As discussed
below, the decays of heavier CP-odd states will be similar to the
situation encountered in the (N)MSSM.

\subsection{CP-even Higgs Boson Decays}
\label{sec:Hdecays}

In the MSSM light CP-even Higgs bosons decay dominantly to $b{\bar b}$
final states. In our calculation we take into account all fermion
final states, including the leading QCD radiative corrections from
\cite{Djouadi:1995gt}. In the SBRP model new decay modes appear, such
as $S^0_i \to JJ$ and, if kinematically allowed $S^0_i \to P^0_jJ$ and $S^0_i
\to P^0_jP^0_k$.  From the latter usually only $S^0_i \to JJ$ has a large
branching ratio (see appendix \ref{sec:non-mssm-decays}).

In Ref.~\cite{Hirsch:2004rw} we have discussed the invisible decays of
the lightest CP-even Higgs boson. We now extend that discussion so as
to include also the next-to-lightest CP-even state which plays an 
important role, if the lightest CP-even state is mainly singlet. 

It is well known that, in contrast to the Standard Model, in the MSSM
(and in the NMSSM) the mass of the lightest CP-even supersymmetric
Higgs boson obeys an upper bound that follows from the D-term origin
of the quartic terms 
in the scalar potential, contained in Eq.~(\ref{scalarpot}). This mass
acquires a contribution from the top-stop quark
exchange~\cite{haber:1991aw,ellis:1991zd,Okada:1990vk}, a fact that
modifies the numerical value of this upper
bound~\cite{haber:1991aw,ellis:1991zd,Okada:1990vk}.  
Many other loops contribute, for a recent two-loop level calculation see,
e.g.~Ref.~\cite{heinemeyer:1998np}.
This limit is slightly relaxed in the NMSSM as opposed to the
MSSM~\cite{ellwanger:1999ji}. 

\begin{figure}[htb]
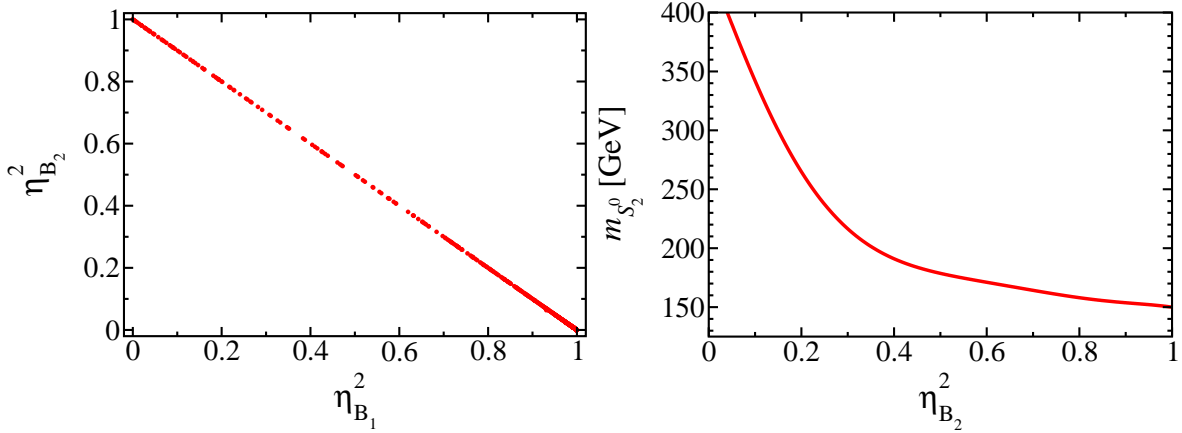

  \centering
  \includegraphics[clip,width=0.47\linewidth]{eta2sq-eta1sq.eps}
  \includegraphics[clip,width=0.47\linewidth]{mH2-eta2sq.eps}
\caption{In the left panel we show the parameter characterizing direct
  production of the second lightest neutral CP-even Higgs boson,
  $\eta_{{\rm B}_2}^2$, as function of the corresponding one for the first 
  lightest neutral CP-even Higgs boson, $\eta_{{\rm B}_1}^2$. To the right:
  Upper limit on the mass of the second lightest CP-even Higgs as a
  function of $\eta_{{\rm B}_2}^2$.}
  \label{fig:eta2-eta1}
\end{figure}

How does this bound emerge in the SBRP model? Since the CP-even sector
contains eight scalars, we cannot diagonalize the corresponding mass
matrices analytically. Therefore we calculate the upper bound on the
Higgs mass numerically, and including the most important radiative
corrections, using formulas from \cite{ellis:1991zd}.  In the SBRP
model it is possible that the lightest CP-even Higgs is mainly a
singlet. However, if this happens, there must exist a light, mainly
doublet Higgs, to which the NMSSM bounds apply. This is shown in Fig.~\ref{fig:eta2-eta1}, 
where we plot (to the left) $\eta_{{\rm B}_2}^2$ as
function of the $\eta_{{\rm B}_1}^2$ and (to the right) the upper limit on
the mass of the second lightest Higgs as function of $\eta_{{\rm B}_2}^2$.
As is seen, if the lightest state is mainly singlet, $\eta_{{\rm B}_1}^2 \simeq 0$, 
therefore $\eta_{{\rm B}_2}^2\simeq 1$, then there is an upper
bound on the second lightest state mass. Vice versa the upper bound applies to the lightest
state if it is mainly doublet.

\begin{figure}[htb]
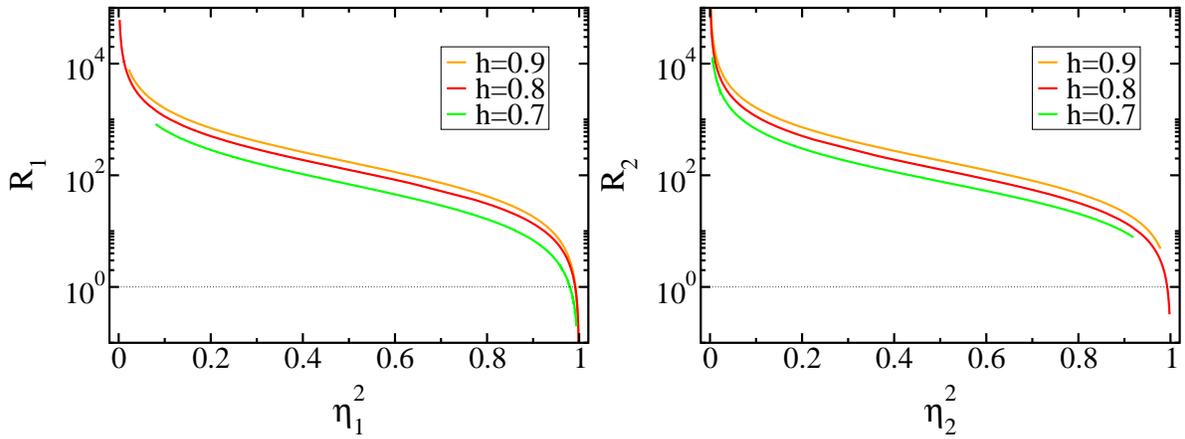

  \centering
  \includegraphics[clip,width=0.47\linewidth]{R1-eta1sq-line.eps}
  \includegraphics[clip,width=0.47\linewidth]{R2-eta2sq-line.eps}
  \caption{To the left (right): Ratio $R_1$ ($R_2$) as a function of 
the direct production parameter, $\eta_{{\rm B}_1}^2$ ($\eta_{{\rm B}_2}^2$), for 
the first (second) lightest neutral CP-even Higgs boson.}
  \label{fig:R-eta}
\end{figure}

As shown previously \cite{Hirsch:2004rw}, one can have large direct
production cross section for the lightest neutral CP-even Higgs boson
as well as a large branching ratio to the invisible final majoron
states. This is demonstrated in the left panel of Fig.~\ref{fig:R-eta} 
for a random but fixed choice of undisplayed parameters. 
We note that a very similar behaviour is also found for the second
lightest state, as seen from the right panel of Fig.~\ref{fig:R-eta}. 
Thus if the lightest state is mainly singlet there
must be a state nearby which is mainly doublet and decays invisibly.

In summary, we have seen that in the SBRP model there is always at
least one light state, which is mainly doublet, and therefore can be 
produced at future colliders. Irrespectively of whether this state is
the lightest or second-lightest Higgs state, it can decay with very
large branching ratio to an invisible final state.

\subsection{CP-odd Higgs Boson Decays}
\label{sec:Adecays}

Light CP-odd Higgs bosons in the MSSM decay according to $P^0_i \to
f{\bar f}$.  The $WW$ channel becomes dominant as soon as kinematically
allowed~\cite{hunters,gunion:1986yn}, however we will not include it
as we are mainly interested in the possibility of invisible decays of
the lowest-lying pseudoscalar.
The formulas for the CP-even and CP-odd Higgs boson MSSM decay 
branching ratios, apart from the larger number of Higgs bosons, are
totally analogous to those of the MSSM~\cite{Djouadi:1995gt}, except
for the prefactors which are determined by the diagonalizing matrices
of our model. The corresponding matrix elements replace the familiar
$\sin (\beta-\alpha)$ and $\cos (\beta-\alpha)$ factors.

In the SBRP we must take into account in addition the decays $P^0_i \to
JJJ$ and, if kinematically allowed, also $P^0_i \to S^0_j J$, $P^0_i \to S^0_j
P^0_k$, $P^0_i \to P^0_j JJ$, $P^0_i \to P^0_j P^0_kJ$ and $P^0_i \to P^0_j P^0_k P^0_m$.
For the lightest Higgs boson we are interested only in $P^0_i
\to JJJ$ and $P^0_i \to S^0_j J$. 
The formulas for the CP-even and CP-odd Higgs boson non-MSSM decay
widths are collected in appendix \ref{sec:non-mssm-decays}.

\subsection{$P^0_i \to S^0_j J$}

The decay width of the CP-odd Higgs boson to a CP-even Higgs boson and
a majoron is given in Eq.~(\ref{eq:widthPSJ}). Using the approximation
Eq.~(\ref{eq:majntl}) we can find the coupling $g'_{ij}$ for the
vertex $S'^0_i P'^0_jJ$ of the majoron with the \textit{unrotated}
neutral scalar $S'^0_i$ and pseudoscalar $P'^0_j$ to leading order in the
small parameter $\frac{v_L}{v}$ as
\begin{eqnarray}
  \label{eq:16}
  g'_{11}&=&\frac{g^2+g'^2}{4}\;\frac{v_d^2v_L^2}{Vv^2}\,,\nonumber\\[+1mm]
  g'_{12}&=&\left(\frac{g^2+g'^2}{4}-h_0^2\right)\frac{v_dv_uv_L^2}{Vv^2}\,,
  \nonumber\\[+1mm]
  g'_{21}&=&-\left(\frac{g^2+g'^2}{4}-h_0^2\right)\frac{v_dv_uv_L^2}{Vv^2}\,,
  \nonumber\\[+1mm]
  g'_{22}&=&-\frac{g^2+g'^2}{4}\;\frac{v_u^2v_L^2}{Vv^2}\,,
  \nonumber\\[+1mm]
  g'_{i1}&=&\frac{-\mu\epsilon_{i-2}}{V}\qquad (i=3,\ldots,5)\,,\nonumber \\[+1mm]
  g'_{i2}&=&\frac{-\epsilon_{i-2}}{V}
            \left(A_{h_{\nu}}+\frac{v_S}{v_R}\hat{M}_R\right)\qquad 
            (i=3,\ldots,5)\,,\nonumber \\[+1mm]
  g'_{61}&=&\frac{v_L^2}{v^2V}
            \left(\sqrt{2}h_0\mu v_d
            -\frac{1}{\sqrt{2}}\left(h_0\hat{M}_{\Phi}+A_{h_0}\right)v_u\right)\,,\nonumber  \\[+1mm]
  g'_{62}&=&\frac{v_L^2}{v^2V}
            \left(-\sqrt{2}h_0\mu v_u
            +\frac{1}{\sqrt{2}}\left(h_0\hat{M}_{\Phi}+A_{h_0}\right)v_d\right)\,,\nonumber  \\[+1mm]
  g'_{71}&=&\frac{-hh_0v_uv_R}{2V}\,,\nonumber \\[+1mm]
  g'_{72}&=&\frac{-hh_0v_dv_R}{2V}\,,\nonumber \\[+1mm]
  g'_{81}&=&\frac{hh_0v_uv_S}{2V}\,,\nonumber \\[+1mm]
  g'_{82}&=&\frac{hh_0v_dv_S}{2V}\,.
\end{eqnarray}
Note that the first four of the above are suppressed by the smallness
of sneutrino vevs, needed to reproduce the observed neutrino
oscillation data.  The coupling $g_{S^0_iP^0_jJ}$ then appears through
mixing, and is given as
\begin{equation}
g_{S^0_iP^0_jJ} = g'_{71} R^{S^0}_{i7}R^{P^0}_{j1} + g'_{72} R^{S^0}_{i7}R^{P^0}_{j2} +
              g'_{81} R^{S^0}_{i8}R^{P^0}_{j1} + g'_{82} R^{S^0}_{i8}R^{P^0}_{j2} \,.
\label{def:gspj}
\end{equation}

\subsection{$P^0_i \to JJJ$}

The decay width of the CP-odd Higgs boson to three majorons is given
in Eq.~(\ref{eq:widthPJJJ}). Using again the approximate equation
giving the profile of the majoron, Eq.~(\ref{eq:majntl}), the coupling
$g'_{i}$ for the vertex $P'^0_i JJJ$ of the majorons with the
\textit{unrotated} neutral pseudoscalar $P'^0_i$, is given as
\begin{eqnarray}
  \label{eq:16y}\nonumber
  g'_1&=&-\frac{3v_L^2}{v^2V^3}h_0 h v_u v_Rv_S\,, \\[+1mm]\nonumber
  g'_2&=&\frac{3v_L^2}{v^2V^3}h_0 h v_d v_Rv_S\,, \\[+1mm]\nonumber
  g'_3 & \sim & g'_4 \sim g'_5\sim {\cal O}(\frac{v_L^3}{v^3})\,,\\[+1mm]\nonumber
  g'_6  & \sim & {\cal O}(\frac{v_L^3\epsilon}{v^3V})\,, \\[+1mm]\nonumber
  g'_7&=&\frac{-3h^2v_Sv_R^2}{V^3}\,, \\[+1mm]
  g'_8&=&\frac{3h^2v_S^2v_R}{V^3}\,.
\end{eqnarray}
Again, the first six of the above vanish in the limit $v_L \to 0$.
Therefore the coupling $g_{P^0_iJJJ}$ for the vertex of the majorons
with the neutral pseudoscalar $P^0_i$ mass eigenstate is
\begin{equation}
g_{P^0_iJJJ}=g'_7 R^{P^0}_{i7}+g'_8 R^{P^0}_{i8}.
\end{equation}

\subsection{Numerical results}

We can see from Eq.~(\ref{eq:16}) that if the CP-odd mass eigenstate
is mainly a Higgs doublet (i.e., its main components are
$P'^0_1=H_d^{0I},\; P'^0_2=H_u^{0I}$ so that its production is not
reduced) then its decays to $S^0_j J$ and $JJJ$ are suppressed as the
corresponding couplings are very small, suppressed by two powers in
$\frac{v_L}{v}$. To find sizeable branching ratios for the decays of
the lightest massive pseudoscalar $P^0_1$, mixing between doublet and singlet
states is therefore required.

\begin{figure}[htbp]
  \centering
  \begin{tabular}{cc}
  \includegraphics[clip,width=0.47\linewidth]{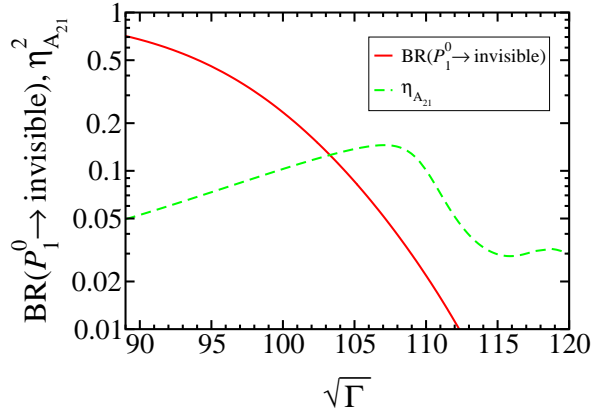}
  \end{tabular}
  \caption{Production cross section (red/solid curve) and invisible
    final states decay branching ratio (green/dashed curve) for the
    lightest CP-odd Higgs boson.}
  \label{fig:brinv}
\end{figure}
As discussed in section \ref{sec:higgs-spectrum}, in order to have
sizeable mixing between doublet and singlet CP-odd Higgs bosons, we
must require that at least one of the singlet states is light, i.e.
the parameter $\Gamma$ should be very roughly of order $\Gamma \sim
\Omega$.  Fig.~\ref{fig:brinv} shows an example. Here, we plot
$\eta^2_{{\rm A}_{21}}$ and ${\rm BR}(P^0_1 \to \hskip1mm {\rm inv})$ as function of
$\sqrt{\Gamma}$ for one fixed, but arbitrary set of other model
parameters. For small values of $\sqrt{\Gamma}$ the lightest massive CP-odd
state is mainly singlet, therefore ${\rm BR}(P^0_1 \to \hskip1mm {\rm inv})$
is close to 1.  However, the production parameter
$\eta^2_{{\rm A}_{21}}$ is small. Increasing $\sqrt{\Gamma}$ increases the
mass of the lightest CP-odd state.  From a certain point onwards, it
is the doublet state which is lightest, compare to Fig.~\ref{fig:masses}. 
This state can have a sizeable production, but the
branching ratio to invisible final states typically is small. Only in
the intermediate region of sizeable mixing between doublet and singlet
states, i.e.  in the region of $\sqrt{\Gamma} \sim 100-115$ GeV of
Fig.~\ref{fig:brinv} one can have both, sizeable production and
sizeable invisible decay.

In summary, the CP-odd Higgs bosons in the SBRP model usually behave
very similar to the situation discussed in the (N)MSSM. However,
sizeable branching ratios to invisible final states are possible when
there are light CP-odd Higgs bosons from both, the doublet and the
singlet sectors.

\section{Discussion}
\label{sec:discussion}

We have carefully analyzed the mass spectra, production and decay
properties of the lightest supersymmetric CP-even and CP-odd Higgs
bosons in models with spontaneously broken R-parity.  We have compared
the resulting mass spectra with what is predicted in the Minimal
Supersymmetric Standard Model, stressing the validity of the upper
bound on the lightest CP-even Higgs boson mass.
We have seen how the presence of the additional scalar singlet states
affects the Higgs production cross sections, both in the Bjorken and
associated modes.

The main difference with respect to the MSSM case comes from the fact
that the spontaneous breaking of lepton number necessarily implies the
existence of the majoron, and this opens new decay channels for
supersymmetric Higgs bosons into ``invisible'' final states.
We have found that the invisible decays of CP-even Higgses can be
dominant, despite the small values of the neutrino masses indicated by
neutrino oscillation data. In contrast, although the decays of the
CP-odd bosons into invisible final states can also be sizeable, this
situation is not generic.

Therefore the existence of invisibly decaying Higgs bosons should be
taken seriosly in the planning of future accelerators, like the LHC
and the ILC. These decays may signal the weak-scale violation of
lepton number in a wide class of theories. Within the supersymmetric
context they are a characteristic feature of the SBRP models. These
can account for the observed pattern of neutrino masses and mixings in
a way which allows the neutrino mixing angles to be cross checked at
high energy accelerators like LHC/ILC. For example, in our model there
is a $b{\bar b}$ plus missing momentum signal associated to the invisible decay of
the lightest CP-even Higgs boson produced in association with a
pseudoscalar. Although this is a standard topology,
also present in the Standard Model and the MSSM, its kinematical
properties in our model differ, as the $JJ$ add up to the CP-even Higgs 
boson mass and $b{\bar b}$ to the CP-odd Higgs 
boson mass.  Further studies to elucidate the impact of these
decay modes for future colliders, should be conducted.  While for the
LHC we may encounter difficulties associated to missing energy
measurements and/or b-tagging, these potential limitations do not
affect in the same way the ILC.

Last, but not least, as already explained, we have restricted our
analysis to Higgs bosons below the $WW$ threshold. Extension to relax
this restriction is totally straightforward, though somewhat less
interesting. Due to the validity of the supersymmetric Higgs boson
mass upper limit we must have one light CP-even Higgs boson which, as
we have shown, is likely to have an important decay into invisible
final states.

\section*{Acknowledgments}

We thank W.~Porod for providing a subroutine for radiative corrections to Higgs masses. 
This work was supported by Spanish grant BFM2002-00345, by the
European Commission Human Potential Program RTN network
MRTN-CT-2004-503369.  M.H. is supported by a MCyT Ramon y Cajal
contract. J.C.R. was supported by the Portuguese \textit{Funda\c{c}\~ao
  para a Ci\^encia e a Tecnologia} under the contract CFTP-Plurianual
and grant POCTI/FNU/50167/2003. A.~V.~M. was supported by Generalitat
Valenciana. 

\appendix
 
\section{Production cross sections}
\label{sec:prod-cross-sect}

In this section we give the formulas for the production cross section
of both channels at an $e^+ e^-$ machine.

\subsection{Bjorken process}

The cross section for the Bjorken process is~\cite{Accomando:1997wt}

\begin{equation}
  \label{eq:10}
\sigma(e^+e^- \to Z^0 S^0_i) = \eta_{{\rm B}_{i}}^2\ 
\frac{G_F^2 M_Z^4}{96 \pi s}\,
\left(v_e^2+a_e^2\right)\, \beta \frac{\beta^2
  +12M_Z^2/s}{(1-M_Z^2/s)^2+\left( \Gamma_Z M_Z/s\right)^2}\,,  
\end{equation}
where 
\begin{equation}
  \label{eq:11}
  v_e=-1+4\sin\theta_W^2, \ a_e=-1, 
\quad \beta=\frac{\lambda(s,M_Z^2,M_{S^0_i}^2)}{s}\ ,
\end{equation}
$\lambda$ is the 2-body phase space function,
\begin{equation}
  \label{eq:12}
  \lambda(a,b,c)=\sqrt{\left(a+b-c\right)^2 -4 a b}
\end{equation}
and the $\eta_{{\rm B}_i}$ are given in Eq.~(\ref{eq:etaB}).

\subsection{Associated production}

The cross section for the associated production is~\cite{pocsik:1981bg}

\begin{equation}
  \label{eq:13}
\sigma(e^+e^- \to S^0_i P^0_j) = \eta_{{\rm A}_{ij}}^2 \ 
\frac{G_F^2 M_Z^4}{96 \pi s}\,
\left(v_e^2+a_e^2\right)\, 
\frac{\beta^3}{(1-M_Z^2/s)^2+\left( \Gamma_Z M_Z/s\right)^2}  
\end{equation}
with
\begin{equation}
  \label{eq:14}
  \beta=\frac{\lambda(s,M_{P^0_j}^2,M_{S^0_i}^2)}{s}
\end{equation}
and the $\eta_{{\rm A}_{ij}}$ are given in Eq.~(\ref{eq:etaA}).

\section{Non-MSSM decays}
\label{sec:non-mssm-decays}

The most characteristic decays of this model which do not exist in the
(N)MSSM are those involving a majoron. In the following we collect the
formulas for these decays.
The most important ones are:

\begin{eqnarray}
\label{eq:widthSJJ}  
&&\Gamma(S^0_i\to  JJ)=\frac{g_{S^0_iJJ}^2}{32\pi m_{S^0_i}}\,,\\[+2mm]
%
\label{eq:widthPSJ}  
&&\Gamma(P^0_i\to S^0_jJ)=\frac{g_{S^0_jP^0_iJ}^2}{16\pi
  m_{P^0_i}^3}\left(m_{P^0_i}^2-m_{S^0_j}^2\right)\,. 
\end{eqnarray}

For completeness we consider also:

\begin{eqnarray}
\label{eq:widthSPJ}  
&&\Gamma(S^0_i\to P^0_jJ)=\frac{g_{S^0_iP^0_jJ}^2}{16\pi
  m_{S^0_i}^3}\left(m_{S^0_i}^2-m_{P^0_j}^2\right)\,, \\[+2mm]
%
\label{eq:widthPSP}  
&&\Gamma(P^0_i\to S^0_jP^0_k)=\frac{g_{S^0_jP^0_iP^0_k}^2}{16\pi
  m_{P^0_i}^3}\lambda(m_{P^0_i}^2,m_{S^0_j}^2,m_{P^0_k}^2)\,, \\[+2mm]
%
\label{eq:widthPJJJ}  
&&\Gamma(P^0_i\to JJJ)=\frac{m_{P^0_i}\, g^2_{P^0_iJJJ}}{3072 \pi^3}\,,\\[+2mm]
%
\label{eq:widthPPJJ}  
&&\Gamma(P^0_i\to P^0_jJJ)=\frac{g_{P^0_iP^0_jJJ}^2}{1024\pi^3
  m_{P^0_i}^3}\left(m_{P^0_i}^4-m_{P^0_j}^4\right)\,, \\[+2mm]
%
\label{eq:widthPPPJ}  
&&\Gamma(P^0_i\to P^0_jP^0_kJ)=\frac{g_{P^0_iP^0_jP^0_kJ}^2}{512\pi^3
  m_{P^0_i}^3}\lambda(m_{P^0_i}^2,m_{P^0_j}^2,m_{P^0_k}^2)\;\times
\left\{ 
  \begin{array}{l}
    \displaystyle
    \frac{1}{2}\, , \quad j=k\cr
    1\, , \quad j \not= k
  \end{array}
\right\}
\nonumber \\[+2mm] 
&&\hskip 5mm \times \, 
\frac{m_{P^0_i}^2\left(m_{P^0_i}^2-2m_{P^0_j}m_{P^0_k}
  \right)+2m_{P^0_j}m_{P^0_k} \left(m_{P^0_j}+m_{P^0_k}\right)^2
  -\left(m_{P^0_j}+m_{P^0_k}\right)^4} {m_{P^0_i}^2-\left(m_{P^0_j}+m_{P^0_k}\right)^2}\,.
%
\end{eqnarray}
The decays $P^0_i \to P^0_jP^0_kP^0_l$ are possible, 
but closed kinematically for the light states of interest,
therefore we do not give here the explicit formulas for the widths.

\bibliographystyle{h-physrev4}

\end{document}